\documentclass[11pt,a4paper,notitlepage,nofootinbib]{revtex4-1}
\usepackage[utf8]{inputenc}
\usepackage{amsmath}
\usepackage{amsfonts}
\usepackage{hyperref}
\hypersetup{
colorlinks=true,
urlcolor=blue,
citecolor=cyan
}

\usepackage{amssymb}
\usepackage{graphicx}
\usepackage[left=2cm,right=2cm,top=2cm,bottom=2cm]{geometry}

\begin{document}
\title{Probing 21cm cosmology and radiative neutrino decays }
\author{Kareem R. H. A. M. Farrag}
\affiliation{School of Physics and Astronomy, Queen Mary University of London, London E1 4NS, UK}
\affiliation{Physics and Astronomy, University of Southampton,
Southampton, SO17 1BJ, U.K}
\begin{abstract}
Here we discuss the recent EDGES observation of a excess in its radio profile and how it can be used to constrain radiative neutrino decays.
\end{abstract}

\maketitle

\begin{center}
 \emph{Presented as a poster at  }
 
{\centering NuPhys2018, Prospects in Neutrino Physics,
Cavendish Conference Centre, London, UK}

{ December 19--21, 2018}
 \end{center}

\section{Introduction}

The EDGES experiment \cite{edgesresults} attempts to probe for the Epoch of Reionisation in its sky-averaged radio spectrum by scanning for absorption spectral lines of primordial hydrogen gas. In particular, the characteristic photons of the 21cm line, with energy $E_{21}=5.87\,\mu \text{eV}$ or $1420 \text{ MHz}$, correspond to the hyperfine splitting of energy levels in neutral hydrogen. This absorption feature is expected to be caused by the radiation emitted from the first stars. Since this happened in the early universe, an observation of hydrogen's spectral lines today will be redshifted due to the universe's cosmological evolution. The authors reported a detection fitting with this search in the frequency range [70 MHz, 90 MHz] corresponding to a redshift $z\sim 17$. Notably, the profile, measured in brightness temperature\footnote{This is the effective temperature measured in EDGES characterising the temperature of the hydrogen gas and the temperature of the photons interacting with it}, appears twice as deep as expected at $-0.5 \text{ K}$, corresponding to a $3.8\sigma$ deviation from predictions within the $\Lambda \text{CDM}$ model. As a function of cosmological parameters and redshift $z$, the brightness temperature $T_{21}(z)$ as measured in EDGES can be expressed as \cite{brightnesstemp}

\begin{equation}
T_{21}(z) \simeq 23 \text{ mK } (1+\delta_b)x_{HI}(z)\left(\frac{\Omega_B h^2}{0.02}\right)\left[\left(\frac{0.15}{\Omega_m h^2}\right)\left(\frac{1+z}{10}\right)\right]^{1/2}\left[1-\frac{T_\gamma(z)}{T_S(z)}\right],
\end{equation}

where $\Omega_B h^2 =0.02226$, $\Omega_m h^2 = 0.1415$ are the baryon and matter abundances and $\delta_B$ is the baryon overdensity, $T_{\gamma}(z)$ is the temperature of the photon bath, and $T_{S}(z)$ the spin temperature of hydrogen. The spin temperature describes the density of states between the ground and excited hydrogen gas, namely
\begin{equation}
\frac{E_{21}}{T_S(z)} = \ln \left(3 \frac{n_0(z)}{n_1(z)}\right),
\end{equation}

where $n_0,n_1$ are the number density for ground and excited hydrogen respectively. The energy $E_{21}$ is the amount required to transition between hydrogen's hyperfine energy levels \footnote{that is the spin configurations of the proton and electron}, where in the ground state the spins of the electron and proton align antiparallel to one another and are aligned parallel in the excited state. Since cosmological parameters are fixed in the $\Lambda$CDM, the only two quantities modifiable in the brightness temperature are $T_\gamma$, the temperature of the CMB at some given redshift, and $T_S$, the spin temperature that characterises the amount of ground state to excited state hydrogen. Alternatively, one might think of modifications to the cosmological model, but here we will simply discuss within the $\Lambda$CDM paradigm.

As mentioned, we have two ways to reproduce the excess in the signal: either cool down the hydrogen to reduce its spin temperature (meaning more ground state hydrogen), so that more 21cm photons might be absorbed (increasing the absorption signal strength); or increase the number of photons through some non thermal production which enhances the brightness temperature measured around the observed frequencies. Indeed, this paradigm is suggested by the ARCADE 2 experiment, which also measured an excess in an analogous search \cite{arcade}. Furthermore, diffuse radio emissions in frequencies of order $O(\text{GHz})$ have been conducted using the ATCA radio array, which can also be recast in the case of nonthermal photon production \cite{atca}.

Many scenarios were explored that involved cooling through some scattering primordial hydrogen off dark matter, some examples can be found in \cite{dmscattering,dmscatteringtwo}. However, these scenarios have since been excluded \cite{millicantexplain,lightdmcantexplain,severedmconstraints} for reasons such as requiring the cross sections to be so small that efficient cooling would not be achieved without some modification to the cosmological model (for example significant deviations to the power spectrum of the Cosmic Microwave Background, CMB for short, from experimental measurement, large scale structure formation and so forth).

\section{Neutrino Radiative decays}
An explanation is the nonthermal production of photons around the 21cm line at the time of the cosmic dawn, through the radiative decay of non-relativistic, quasi-degenerate mass neutrinos. Now in the case of active-to-active radiative decays this would require neutrinos heavier than the Planck bound \cite{planckmassconstraint} and a nonzero effective magnetic moment for the active neutrinos, which is also tightly constrained\footnote{ In upcoming experiments the effective magnetic moment for active neutrinos will be constrained down to order $O(10^{-12})\mu_B$, where $\mu_B$ is the Bohr magneton \cite{numagmomentfuture}} \cite{numagmomentsone}.

If one however considers the decay between active and sterile neutrinos $\nu_i \rightarrow \nu_s \gamma$, then any spectral distortions to the CMB can be avoided resulting from neutrino-photon couplings and a nonthermal component of photons in the extragalactic spectrum might be produced which explains the EDGES excess. As shown in \cite{ourpaper}, one expects that given such active-to-sterile neutrino radiative decays, there are two possible solutions that could explain the EDGES, ARCADE 2 or ATCA excesses. Either, the non-relativistic neutrinos have aleady decayed and produced a nonthermal contribution to the spectrum; else, the lifetime of the neutrinos is longer than the lifetime of the universe, and hence observation of these newly formed photons modifies the spectrum today, as shown in Figure 1. Note that through these excesses we are able to constrain the lifetime and masses of these light neutrinos.

\begin{figure}[h!]
\centering
\includegraphics[scale=0.35]{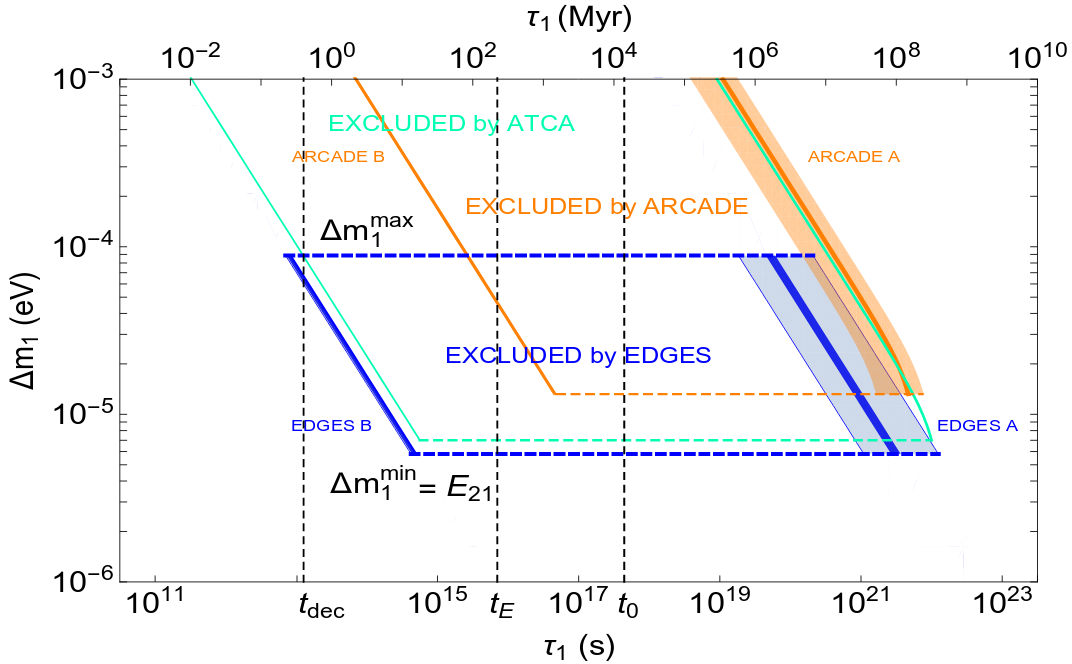} 
\caption{This plot from \cite{ourpaper} shows the exclusion regions for the EDGES, ARCADE2 and ATCA experiments. The allowed mass ranges and lifetimes for the neutrinos lie along the diagonal lines assuming each experiment's profile is correct, with the shaded regions indicating error from the relevant measurements. Note there are upper and lower bounds on each profile. The lower bound comes from the requirement of providing the minimum amount of energy to produce a 21cm photon. The upper bound results from the constraint that the neutrinos are non-relativistic at the time of their decay.}
\end{figure}

\section{Conclusion}
We require more observations to test if the EDGES anomaly holds, but it has shown that the result can probe the process of neutrino radiative decays via the production of 21cm photons. Experiments can and will be able to search the 21cm line and test the report by the EDGES experiment, such as the SARAS 2 spectral radiometer \cite{saras},  and LEDA
experiment \cite{leda}. If indeed EDGES is substantiated by other experiments and confirmed, precision measurements of the absorption spectrum as a function of redshift
could potentially be used to more rigorously test neutrino radiative decays.  Since photons couple to charge, one might think of both standard and beyond the standard model scenarios that generate the radiative process. Further to this, the current scenario assumes non-relativistic decays; of course, allowing for a momentum distribution for the neutrinos would realise a more complete treatment that could further test different models generating the radiative process.

\vspace{1em}

{\textbf{Acknowledgements}}

KF acknowledges financial support from the NExT/SEPnet Institute.

\end{document}